\documentclass[1p]{elsarticle}

\usepackage{hyperref}
\usepackage{subfig}
\usepackage{amsmath}
\usepackage{booktabs}
\usepackage{graphicx}
\usepackage{color}

\DeclareGraphicsExtensions{.pdf}

\journal{Simulation Modelling Practice and Theory}

\begin{document}

\begin{frontmatter}

\title{New Trends in Parallel and Distributed Simulation:\\ from 
    Many-Cores to Cloud Computing
\footnote{The publisher version of this paper is available at \url{http://dx.doi.org/10.1016/j.simpat.2014.06.007}. \textbf{{\color{red}Please cite as: Gabriele D'Angelo, Moreno Marzolla. New trends in parallel and distributed simulation: From many-cores to Cloud Computing. Simulation Modelling Practice and Theory, Elsevier, vol. 49 (December 2014)}}.
An early version of this work appeared in~\cite{gda-hpcs-11}. This paper is an extensively revised and extended version in which significantly more than 30\% is new material.}}

\author{Gabriele D'Angelo\corref{cor1}}
\ead{g.dangelo@unibo.it}

\author{Moreno Marzolla}
\ead{moreno.marzolla@unibo.it}

\address{Department of Computer Science and Engineering. University of Bologna, Italy.}

\cortext[cor1]{Corresponding Author. Address: Department of Computer Science and Engineering. University of Bologna. Mura Anteo Zamboni 7. I-40127, Bologna. Italy. Phone +39 051 2094511, Fax +39 051 2094510} 

\begin{abstract}
  Recent advances in computing architectures and networking are
  bringing parallel computing systems to the masses so
  increasing the number of potential users of these kinds of systems.
  In particular, two important technological evolutions are happening at the
  ends of the computing spectrum: at the ``small'' scale,
  processors now include an increasing number of independent execution
  units (cores), at the point that a mere CPU can be considered a
  parallel shared-memory computer; at the ``large'' scale, the Cloud
  Computing paradigm allows applications to scale by offering
  resources from a large pool on a pay-as-you-go model.  Multi-core
  processors and Clouds both require applications to be
  suitably modified to take advantage of the features they provide.
  Despite laying at the extreme of the computing architecture
  spectrum -- multi-core processors being at the small scale, and Clouds
  being at the large scale -- they share an important common trait: both are
  specific forms of parallel/distributed architectures. As such, they
  present to the developers well known problems of synchronization,
  communication, workload distribution, and so on. Is parallel and
  distributed simulation ready for these challenges? In this paper,
  we analyze the state of the art of parallel and distributed
  simulation techniques, and assess their applicability to multi-core
  architectures or Clouds. It turns out that most of the current
  approaches exhibit limitations in terms of usability and adaptivity
  which may hinder their application to these new computing
  architectures. We propose an adaptive simulation mechanism, based
  on the multi-agent system paradigm, to partially address some of those 
  limitations. While it is unlikely that a
  single approach will work well on both settings above, we argue that
  the proposed adaptive mechanism has useful features which make it attractive
  both in a multi-core processor and in a Cloud system. These features
  include the ability to reduce communication costs by migrating
  simulation components, and the support for adding (or removing) nodes
  to the execution architecture at runtime. We will also show that,
  with the help of an additional support layer, parallel and distributed
  simulations can be executed on top of unreliable resources.
\end{abstract}


\end{frontmatter}

\section{Introduction}\label{sec:introduction}

In the last decade, computing architectures were subject to a
significant evolution. Improvements happened across the whole spectrum
of architectures, from individual processors to geographically
distributed systems.

In 2002, Intel introduced the Hyper-threading (HT) technology in its
processors~\cite{HT}. An HT-enabled CPU has a single execution unit but
can store two architecture states at the same time. HT processors
appear as two ``logical'' CPUs, so that the Operating System can
schedule two independent execution threads. This allows a marginal
increase of execution speed due to the more efficient use of the
shared execution unit. Improvements of miniaturization and chip
manufacturing technologies naturally led to the next step, where
multiple execution units (``cores'') are put on the same processor
die. Multi-core CPUs can actually execute multiple instructions at the
same time. These types of CPUs are now quite ubiquitous, being found
on devices ranging from high-end servers to smartphones and tablets;
CPUs with tens or hundreds of processors are already
available~\cite{tile64}. The current generation of desktop processors
often combines a multi-core design with HT. This means that each
physical core will be seen by the Operating System as two virtual
processors.

While many multi-core processors are homogeneous shared-memory
architectures, meaning that all cores are identical, asymmetric or
heterogeneous multi-core systems also exist. A notable example is the
Cell Broadband Engine~\cite{cell}, that contains two PowerPC cores and
additional vector units called Synergistic Processing Elements
(SPE). Each SPE is a programmable vector co-processor with a separate
instruction set and local memory for instructions and data. Other
widely used asymmetric multi-core systems include General-Purpose
Graphics Processing Units (GP-GPU)~\cite{gpu}, which are massively parallel
devices containing up to thousands of simple execution units connected
to a shared memory. The memory is organized in a complex hierarchy,
since RAM chips do not provide enough bandwidth to feed all cores at
maximum speed. GP-GPUs are generally implemented as add-on cards which
are inserted into a host computer. The host CPU then sends data and
computation to the GPU through data transfers from host to device
memory; results are fetched back to host memory when computation
completes.

In modern multi-core processors, the level of complexity introduced by
the hardware is quite high; unfortunately, this complexity can not be
ignored, and instead it has to be carefully exploited at the
application level. In fact, writing efficient applications for
multi/many-core processors requires a detailed knowledge of the
processor internals to orchestrate communication and computation
across the available cores. 

Not only hardware is undergoing the major changes just described: at
the software layer the Everything as a Service (EaaS) is gaining 
momentum thanks to Cloud Computing. According to NIST~\cite{mell11}, 
Cloud Computing is a model for enabling on-demand network access to 
a shared pool of resources.

Depending on the type of resource which is provided, we can identify
different Cloud service models. In a Software as a Service (SaaS)
Cloud, customer receives access to application services running in the
Cloud infrastructure;``Google Apps'' is an example of a SaaS Cloud.  A
Platform as a Service (PaaS) Cloud provides programming languages,
tools and a hosting environment for applications developed by the
Cloud customer. Examples of PaaS solutions are AppEngine by Google,
Force.com from SalesForce, Microsoft's Azure and Amazon's Elastic
Beanstalk. Finally, an Infrastructure as a Service (IaaS) Cloud
provides its customers with low level computing capabilities such as
processing, storage and networks where the customer can run arbitrary
software, including operating systems and applications.

The architectural evolutions described above are not completely
transparent to existing applications, which therefore should be modified 
to take full advantage of new features. In
this paper, we focus on a specific class of software applications,
namely simulation tools. Simulation can be very demanding in terms of
computational resources, for example when dealing with large and
detailed models~\cite{FUJ00}. 
Parallel and Distributed Simulation
(PADS)~\cite{FUJ00} aims at studying methodologies
and techniques for defining and executing simulation models on
parallel and distributed computing architectures. PADS can be
considered a specialized sub-field of parallel computing and
simulation; as such, it shares many common issues with other types of
parallel application domains (e.g., load balancing, partitioning,
optimizing communication/computation ratio, and so on). 

After a brief introduction on PADS, we review the most important
challenges faced by PADS on multi-core processors and Clouds, and 
argue that it needs ad-hoc solutions which are quite different from 
those commonly used to develop parallel or distributed applications. 
We see that, the most common PADS technologies are not adequate for 
coping with such architectural evolution. In particular, we analyze 
the limitations of current PADS approaches in terms of performance, 
functionality and usability. Moreover, we discuss the problem of 
finding new metrics for the evaluation of the PADS performance. 
In fact, in our view, such metrics have to consider many different
aspects such as the execution speed, the cost of resources and the
simulation reliability. Finally, we suppose that an adaptive 
approach can overcome many of these problems and therefore we 
describe a new parallel/distributed simulation technique that is
based on the dynamic partitioning of the simulation model.

This paper is organized as follows. In Section~\ref{sec:background} we
give some background notions on Parallel and Distributed Simulation.
Section~\ref{sec:challenges} deals with the challenges presented to
PADS by many-core architectures and Clouds. In
Section~\ref{sec:artisgaia} we describe and discuss our proposal aimed
to obtain more adaptable PADS. Finally, concluding remarks will be
discussed in Section~\ref{sec:conc}.

\section{Background}\label{sec:background}

A computer simulation is a software program that models the evolution
of some real or abstract system over time. Simulation can be useful to
evaluate systems before they are built, to analyze the impact of
changes on existing systems without the need to physically apply the
changes, or to explore different design alternatives.

In this paper, we focus on Discrete Event Simulation
(DES)~\cite{Law:1999:SMA:554952}: in a DES, the model evolution
happens at discrete points in time by means of \emph{simulation
  events}. Hence, the simulation model is evolved through the
creation, delivery and execution of events. In its simplest form, a
DES is implemented using a set of state variables, an event list, and
a global clock representing the current simulation
time~\cite{Law:1999:SMA:554952}.

A simulator capable of exploiting a single execution unit (e.g.,~a
single CPU core) is called a \emph{sequential} (monolithic)
simulator. In a sequential simulation the execution unit processes
events in non-decreasing timestamp order, and updates the model state
variables and the global clock accordingly. The main advantage of this
approach is its simplicity, but is clearly inappropriate for large and
complex models. First, complex models might generate a large number of
events, putting a significant workload on the CPU. Furthermore, the
memory space required to hold all state information may exceed the RAM
available on a single processor~\cite{1668384}.

Parallel and Distributed Simulation (PADS) relies on partitioning the
simulation model across multiple execution units. Each execution unit
manages only a part of the model; in PADS each execution unit handles
its local event list, but locally generated events may need to be
delivered to remote execution units. Partitioning of the simulation
model allows each processor to handle a portion of the state space and
a fraction of the events; the parallel execution of concurrent events
can also lead to a reduction of the simulation run time. Additional
advantages of PADS include: the possibility to integrate simulators
that are geographically distributed, the possibility to integrate a
set of commercial off-the-shelf simulators, and to compose different
simulation models in a single simulator~\cite{FUJ00}.

A PADS is obtained through the interconnection of a set of model
components, usually called Logical Processes (LPs). Therefore, each LP
is responsible for the management of the evolution of a subset of the
system and interacts with the other LPs for all the synchronization
and data distribution operations~\cite{FUJ00}. In practice, each LP is
usually executed by a processor core. 

The scientific literature distinguishes between \emph{parallel} and
\emph{distributed} simulation although the difference is quite
elusive. Usually, the term \emph{parallel simulation} refers to
simulation modeling techniques using shared-memory multiprocessors or
tightly coupled parallel machines. Conversely, \emph{distributed
  simulation} refers to simulation techniques using loosely coupled
(e.g.,~distributed memory) architectures~\cite{perumalla2007}. It
should be observed that, in practice, execution platforms are often
hybrid architectures with a number of shared-memory multiprocessors
connected through a LAN.

The lack of a global state and the presence of a network that
interconnects the different parts of the simulator has some important
consequences:

\begin{itemize}

\item the simulation model must be partitioned in components (the
  LPs)~\cite{bagrodia98}. In some cases, the partitioning is guided by
  the structure and the semantics of the simulated system. For
  example, if the simulation involves groups of objects such that most
  interactions happen among objects of the same group, then it is
  natural to partition the model by assigning all the objects in the
  same group to the same LP.  However, in the general case, it may be
  difficult if not impossible to identify an easy way to partition the
  model. As a general rule, partitioning should guarantee that the
  workload is balanced across the available LPs, and the communication
  between the LPs is
  reduced~\cite{5356113,761559,4262808,gda-ijspm-2009}.  Therefore,
  the partitioning criteria strictly depends both on the simulation
  model to be partitioned~\cite{685269}, on the underlying hardware
  architecture on which the model will be executed, and the
  characteristics of the synchronization algorithm that is
  implemented~\cite{Boukerche:1997:DLB:268826.268897};

\item the results of a parallel/distributed simulation must be
  identical to those of a sequential execution. This means that
  \emph{causal consistency}~\cite{Lamport1978,FUJ00} among simulation
  events must be guaranteed (see Section~\ref{Synchronization}).
  While causal consistency can be trivially achieved in sequential
  simulations by simply executing events in non-decreasing timestamp
  order, it is much harder to achieve in PADS, especially on
  distributed-memory systems where LPs do not share a global view of
  the model. Therefore, PADS must resort to some kind of
  \emph{synchronization} among the different parts that compose the
  simulator. Specific algorithms are needed for the synchronization of
  the LPs involved in the execution process;

\item each component of the simulator will produce state updates that
  are possibly relevant for other components. The distribution of such
  updates in the execution architecture is called \emph{data
    distribution}. For obvious reasons, sending all state updates to
  all LPs is impractical. A better approach is to match the data
  production and consumption using some publish-subscribe method in
  which LPs can declare which type of data they are interested in and
  what they produce~\cite{Jun:2002:ESM:564062.564074}.

\end{itemize}

Partitioning, synchronization and data distribution are important
problems in the context of PADS. Synchronization is of particular
interest since parallel and distributed simulations tend to be
communication bound rather than computation bound. Therefore, the
choice of the synchronization algorithm, as well as the type of communication
network used by the execution host, plays the most important role in
determining the performance and scalability of the simulation model.

\subsection{Synchronization}\label{Synchronization}

Implementing a parallel simulation requires that all simulation events are delivered
following a message-based approach. Two events are said to be in
\emph{causal order} if one of them may depend on the
other~\cite{Lamport1978}. Enforcing causal order in a PADS requires
that the different LPs are synchronized. That is because every LP can
proceed at different speed, and therefore each partition of the
simulation model may evolve at a different rate. Different solutions
to this problem have been proposed, which can be broadly summarized in
three main families: \emph{time-stepped}, \emph{optimistic} and
\emph{conservative} simulations.

In a \emph{time-stepped synchronization}, the simulated time is
divided in fixed-size intervals called \emph{timesteps}. All LPs are
required to complete the current timestep before moving to the next
one~\cite{1261535}. The implementation of this approach requires a
barrier synchronization between all LPs at the end of each step. 
Time-stepped synchronization is quite easy to implement, and it is most
appropriate for models which already exhibit some form of lockstep
synchronization (e.g., VLSI circuit simulation, where the clock signal
is used to synchronize the functional blocks of the circuit). For
other applications, it may be difficult to define the correct timestep
size; furthermore, time-stepped mechanisms require that all newly
generated events must be scheduled for future timesteps only (i.e., it
is not possible to generate a new event to be executed during the
current timestep), and this requirement can be too limiting for some
applications.

The goal of the \emph{conservative synchronization} approach is to
prevent causality violations while allowing the global simulation time
to advance at an arbitrary rate. Therefore, a LP can process an event
with timestamp $t$ only if no other event with timestamp $t' < t$ will
be received in the future. The Chandy-Misra-Bryant
(CMB)~\cite{misra86} algorithm can be used to guarantee causal
delivery of simulation events in a distributed system. The CMB
algorithm requires that each LP $i$ has a separate incoming message
queue $Q_{ji}$ for each LP $j$ it receives events from. Each LP is
required to generate events in non-decreasing timestamp order, so that
LP $i$ can identify the next ``safe'' event to process by simply
checking all incoming queues $Q_{ji}$: if all queues are nonempty,
then the event with lowest timestamp $t$ is safe and can be processed.
If there are empty queues, this mechanism can lead to deadlock, which
can be avoided by introducing null messages with no semantic
content. The goal of these messages is to break the circular chain
that is a necessary condition for deadlock. The main drawback of the
CMB synchronization mechanism is the high overhead introduced by null
messages in terms of both network load and computational overhead.

The \emph{optimistic synchronization} approach does not impose the
execution of safe messages only: each LP can process the events as
soon as they are received. Causality violations can happen if some
message (called a \emph{straggler}) with timestamp in the past is
received by some LP. When a causality violation is detected, the
affected LP must roll back its local state to a simulation time prior
to the straggler timestamp. The roll back mechanism must be propagated
to all other LPs whose simulation time has advanced past that of the
straggler~\cite{timewarp,quaglia2003}. This may trigger a rollback
cascade that brings back the whole simulation to a previous state,
from where it can be re-executed and process the straggler in the
appropriate order.
Optimistic synchronization mechanisms require that each LP maintains
enough state data and a log of sent messages, in order to be able to
perform rollbacks and propagate them to other affected LPs. A suitable
snapshot mechanism must be executed periodically in order to compute a
global state that is ``safe'', meaning that it can not be rolled back.
This is essential to reclaim memory used for events and state
variables that become no longer necessary. This problem is known as
the \emph{Global Virtual Time} (GVT) computation, where the GVT
denotes the earliest timestamp of an unprocessed message in an
optimistic simulation. Computing the GVT is not unlikely taking a
snapshot of a distributed computation, for which the well known
Chandy-Lamport algorithm~\cite{ChandyL85} can be used. However, in the
context of PADS, more efficient ad-hoc algorithms have been
proposed~\cite{gvt95}.

\subsection{Software tools}

Many tools have been developed to support the implementation of PADS,
some of which are described below.

$\mu$sik~\cite{Perumalla:2005:MPS:1069810.1070161} is a multi-platform
micro-kernel for the implementation of parallel and distributed
simulations. The micro-kernel provides a rich set of advanced features
such as the support for reverse computation and some kind of
automated-load balancing. The last version of $\mu$sik was released in
2005 and now the development seems to have stopped.
SPEEDES~\cite{Steinman:2003:SPF:824475.825880} and the WarpIV
Kernel~\cite{Steinman_08s-siw-025warpiv} have been used as testbeds
for investigating new technologies, such as the Qheap, a new data
structure for event list management. Furthermore, SPEEDES has been
used for many seminal works on load-balancing in optimistic
synchronization~\cite{Wilson:1998:ELM:293172.293267,Wilson:1995:ALB:224401.224691}. Finally,
PRIME~\cite{prime} and PrimoGENI~\cite{5936747} have specific focus on
very high scalability and real-time constraints, mainly in complex
networking environments.

One important advance in the field of PADS is the IEEE 1516--High
Level Architecture (HLA) standard~\cite{ieee1516} that has been
approved in 2000 and recently revised under the SISO HLA-Evolved
Product Development Group~\cite{hla-evolved}. HLA is a standard
architecture for distributed simulation that allows computer
simulations to interact with other simulations using standard
interfaces.

The interest in IEEE 1516 has been initially very strong and it is
still fundamental for interoperability and reusability of simulators.
Despite of this, some drawbacks have been reported: the software 
architecture required by HLA~\cite{Davis:1999:HLA:324898.325337},
the lack of interoperability among different 
implementations~\cite{hlainteroperability}, its complexity and steep
learning curve~\cite{4736178} and, in some cases, poor execution
time~\cite{caihlaperformance}.

\begin{table}[t]
\centering\begin{small}\begin{tabular}{lllll}
\toprule
{\bf Name} & {\bf Author} & {\bf Bindings} & {\bf Compliance} & {\bf License} \\
\midrule
RTI NG Pro~\cite{rti-ng-pro} & Raytheon & C++, Java & Full &Commercial \\
FDK~\cite{fdk} & Georgia Tech & C/C++ & Partial & Non-free\\
MAK RTI~\cite{mak} & VT MAK & C/C++, Java & Full & Commercial \\
Pitch pRTI~\cite{pitch} & Pitch Technologies & Multiple & Full & Commercial \\
CERTI~\cite{certi} & ONERA & Multiple & Partial & GPL/LGPL \\
OpenSkies RTI~\cite{openskies} & Cybernet Systems & C++ & Partial & Commercial \\
Chronos RTI~\cite{chronos} & Magnetar Games & C++/.NET & Unknown & Commercial \\
The Portico Project~\cite{portico} & - & C++, Java & Partial & CDDL \\
SimWare RTI~\cite{simware} & Nextel Aerospace & C++ & Partial & Commercial \\
Open HLA~\cite{openhla} & - & Java & Partial & Apache\\
OpenRTI~\cite{openrti} & FlightGear & C++,Python & Unknown & LGPL\\
\bottomrule
\end{tabular}\end{small}
\caption{Some implementations of the HLA specification.}\label{tab:hla}
\end{table}

Table~\ref{tab:hla} shows some of the available (full or partial)
implementations of the HLA specification. In general, as expected,
commercial proprietary implementations are more compliant with the
specification and feature-rich. 
However, the Portico Project
implementation is very promising even if it still does not support the
new HLA-evolved interface specification.

\subsection{PADS on Many-core and Cloud Computing Platforms}

The usage of many-core chips for running PADS is not a novel topic.
Most of the existing works deal with the usage of optimistic Time
Warp-based tools and their performance
evaluation~\cite{simul2011,Child:2012:DAC:2372596.2372609,6267855}.
For example, in~\cite{simul2011} it is shown that the architecture of
many-core processors is rather complex and that some techniques are
needed for increasing the execution speed (e.g.,~multi-level Time
Warp, multi-level memory-aware algorithms, detailed frequency control
of cores). The opportunity to control the speed of each CPU core is
further investigated in~\cite{Child:2012:DAC:2372596.2372609}. In this
case, the idea is that by controlling the execution speed of each
core, it is possible to limit the number of roll-backs in optimistic
synchronization. This approach is not different from adjusting the
speed of CPUs, a widely investigated topic, but in this case the
adjustment is done on different regions of the same
chip. In~\cite{6267855} a multi-threaded implementation of an
optimistic simulator is studied on a 48 core computing platform. This
implementation, avoiding multiple message copying and reducing the
communication latency, is able to reduce the synchronization overhead
and improve performance. Finally, \cite{HuiweiLv:2010:PPC:2015554.2015814} 
shows how the Godson-T Architecture Simulator (GAS) has been ported to
a many-core architecture with very good performance results but this has 
been possible mainly thanks to the loosely coupled nature of events in the
specific simulation model of the GAS.\\
In~\cite{MunckVB14} the authors evaluate the performance of a
conservative simulator on a multi-core processor. Since the performance
of conservative synchronization protocols can be affected by the large
number of null messages generated to avoid deadlock conditions, the
authors evaluate a number of known optimizations to reduce this
overhead. A new hybrid protocol is finally proposed in order to
combine the effectiveness of existing solutions.\\

A growing number of papers address the problem of running PADS on
Cloud execution architectures. The early works on this topic have
mainly considered the usage of Private
Clouds~\cite{fuj2010-cloud1,fuj2010-cloud2,5564701};
more recent works deal with the usage of Public Clouds and the related
problems~\cite{Vanmechelen2013126,Yoginath:2013:EEC:2486092.2486118,Aizstrauts:2012:ECE:2230596.2230628}.
In~\cite{Liu2012,Liu:2012:CSS:2372596.2372613} the authors describe
the state of the art of PADS on Cloud and propose a specific
architecture based on a two-tier process partitioning for workload
consolidation. In~\cite{Li:2013:AOH:2486092.2486119} an optimistic
HLA-based simulation is considered and a mechanism for advancing the
execution speed of federates at comparable speed is
proposed. In~\cite{6245700}, an approach based on concurrent
simulation (that we have investigated in the
past~\cite{gda-pads-2005}) has been ported to the Cloud.  Finally, it
is worth to mention~\cite{Mancini:2012:SCU:2310096.2310201} in which
the authors propose to build simulations on the Cloud using handheld
devices and a web-based simulation approach.\\

Even if many-core CPUs and Cloud Computing environments are very different
execution architectures, with specific problems, it should be 
clear that the state of the art is, in both cases, made of quite
specific solutions. Solutions that are tailored for a specific 
problem or a given simulation mechanism. In all cases with its 
own benefits and drawbacks.

\section{Challenges}\label{sec:challenges}

As already observed in the introduction, recent technological
evolutions of computing systems require software developers to adapt
their applications to new computing paradigms. Simulation developers,
in particular those working on PADS, are no exception. In this section
we discuss more specifically how multi-core processors and Cloud
Computing affect simulation users. 

The following issues that needs to be addressed by future PADS
systems:

\begin{enumerate}
\item \textbf{Transparency}: parallel simulation systems should provide
  the possibility to hide low level details (e.g., synchronization and
  state-management issues) to the modeler, should the modeler be
  unwilling to deal manually with these issues.
\item \textbf{Simulation as a Service}: what are the main challenges
  that should be addressed in order to provide simulation ``as a
  service'' using a Cloud?
\item \textbf{Cost Models}: Cloud resources are usually provisioned on
  a pay-per-use pricing model, with ``better'' resources (e.g., more
  powerful virtual processors, more memory) having higher cost. Is
  this pricing model suitable for PADS?
\item \textbf{Performance}: Cloud Computing allows applications to
  dynamically shape the underlying computing infrastructure by
  requesting resources to be added/removed at run time. Cloud-enabled
  PADS systems should be extended to make use of this opportunity.
\end{enumerate}

\subsection{Transparency}\label{sec:challenges.transparency}

Two of the main goals of the research work on PADS in the last decades
were: i) \emph{make it fast}; ii) \emph{make it easy to
  use}~\cite{fujtut2000}. Today, PADS can be very fast when executed
in the right conditions~\cite{Perumalla:2007:STW:1242531.1242543},
e.g., when the simulation model is properly partitioned, the
appropriate synchronization algorithm is used and the execution
architecture is fast, reliable and possibly homogeneous. In terms of
usability, however, PADS does not yet work ``out of the box''.

In principle, the user of simulation tools should focus on modeling
and analysis of the results. In practice, the choice of a specific
tool (e.g., a sequential simulator of some kind) will preclude a
future easy migration of the model towards another sequential
simulator of a different kind, or towards a PADS system.  For example,
if a user wishes to migrate from a sequential to a parallel
simulation, he will likely need to cope with such details as the
choice of synchronization algorithm (optimistic or conservative), the
allocation of simulated entities on physical execution units, details
of the messaging system that allows simulated entities to communicate,
and so on.

Specifically, in PADS it is necessary to partition the simulation
model across the available execution units. A good partitioning
strategy requires that the communication among the partitions are
minimized (cluster interactions) and the workload is evenly
distributed across the execution units (load balancing).
In case of static simulation models, i.e., models where the
communication patterns among partitions do not change significantly
over time, it is possible to statically define the partitions at model
definition time. It should be observed that, even in the static
scenario, identifying an optimal simulation model partitioning where
inter-partition communication is minimized is a known NP-complete
problem~\cite{Garey1979}.  If the communication pattern changes over
time, then some form of adaptive load balancing should be
employed~\cite{par-sim-today}. Adaptive load balancing is appealing
since it can be applied also to static models to allow the ``best''
partitioning to be automatically identified without any specific
application-level knowledge. Adaptive load balancing is difficult to
implement (see~\cite{Kurve2011} for an heuristic approach based on
game theory), and is still subject of active research

A better way to deal with the problems above would be to separate the
simulation model from the underlying implementation mechanisms, which
in the case of PADS means that low level details such as
synchronization, data distribution, partitioning, load balancing and
so on should be hidden to the user. This proved to be quite difficult
to achieve in practice: an example is again the High Level
Architecture (IEEE 1516)~\cite{ieee1516} that supports optimistic
synchronization, but the low-level implementation of all the support
mechanisms (such as rollbacks, see Section~\ref{sec:background}) is
left to the simulation modeler~\cite{Santoro2006}. Therefore, an existing 
conservative HLA simulation can hardly be ported to an optimistic 
simulation engine.

\subsection{Simulation as a Service}

As already described in Section~\ref{sec:background}, some work on
Cloud-based PADS has already been
done~\cite{fuj2010-cloud1,fuj2010-cloud2,5564701}. In many of these
works, Cloud technology is used for executing simulations in private
Clouds. The goal is to enjoy the typical benefits of Cloud Computing
(elasticity, scalability, workload consolidation) without giving up
the guarantees offered by an execution environment over which the user
has a high degree of control. However, the vast majority of simulation
users do not own a Private Cloud, therefore it would be much more
interesting to run parallel simulations on a Public Cloud, using
resources rented with a pay-per-usage model.

Obviously, the partitioning problem described above still holds;
additionally, due to its nature, a Public Cloud provides a somewhat
less predictable environment to applications. For example, execution
nodes provided to users may be located in different data centers, have
slightly different raw processing power, and be based on physical
resources (e.g., processors) that are shared with other Cloud
customers through virtualization techniques.

Considering the non-technical aspects, one of the potential advantages
of the Public Cloud is the existence of multiple competing providers,
that could -- at least in principle -- lead to a a ``market of
services''. In other words, the same service (e.g.,~computation) could
be provided by many vendors and therefore the price is the result of a
market economy of supply and demand. 
To take advantage of the competition among vendors, and the resulting
price differences, the user must be able to compose services from
different providers. The result is an heterogeneous architecture where
performance and reliability aspects must be carefully taken into
account.

However, the scenario above is unlikely, due to the lack of
interoperability and the use of different APIs by different
vendors. While in the short term vendor lock-in is beneficial to
service providers, there already is pressure towards the use of open
standards in Cloud Computing, especially in the academic/research
community, culminated in the Open Cloud Computing Interface (OCCI)
specifications~\cite{occi}. It remains to be seen if and when open
Cloud standards will be able to carve into the corporate community.

\subsection{Cost Models}\label{sec:cost-model}

The performance of a simulator is usually evaluated on the basis of
the time needed to complete a simulation run (Wall-Clock-Time,
WCT). However, this is not the only metric, and in some scenario it
may also not be the most relevant one.  For example, if the simulation
is going to be executed on resources acquired from a Cloud provider,
another important metric is the total cost of running the simulation,
which is related both to to the WCT and also to the amount of
computation and storage resources acquired. Therefore, the user of
simulation tools will have to decide:
\begin{itemize}
\item How much time he can wait for the results;
\item How much he wants to pay for running the simulation.
\end{itemize}

The first constraint (how much time the user is willing to
  wait) should be always taken into consideration in any cost model.
  If no upper bound is set on the total execution time, then the cost
  of running the simulation essentially drops to zero, since that
  would allow the use of extremely cheap (if not totally free)
  resources of proportionally low performance. In practice, nobody is
  really willing to wait an arbitrary long time; hence a maximum
  waiting time is always defined, and this influences the cost of the
  resources that must be allocated to run the simulation: faster (and
  therefore expensive) resources are needed if the simulation result
  has to be provided quickly; slower (cheaper) resources can be
  employed if the user can tolerate a longer waiting time.

It is clear that, in a Public Cloud scenario, every computation and
communication overhead should be minimized or at least carefully
scrutinized. However, the two traditional synchronization mechanisms
used in PADS (the Chandy-Misra-Bryant algorithm and the Time Warp
protocol) have not been developed with these considerations in mind.

The Chandy-Misra-Bryant (CMB) algorithm~\cite{misra86} is one of the most
well-known mechanisms used for conservative synchronization.  Due to
its nature, it requires the definition of some artificial events with 
the aim of making the simulation
proceed. The number of such messages introduced by the synchronization
algorithm can be very large~\cite{devries1990,rizvi2006}. Obviously, 
this communication overhead has a big effect on the WCT. Over the
years, many variants have been proposed to reduce the number of such
messages~\cite{Su:1988:VCD:865909} and therefore reduce the 
communication cost. Despite of this, in some cases, the CMB 
synchronization can still be prohibitive under the performance 
viewpoint. On the other hand, the consideration that computation is
much faster and cheaper than communication is at the basis of optimistic
synchronization, e.g., the Time Warp protocol~\cite{timewarp}.  

Both the conservative and optimistic synchronization mechanisms
described above are not well suited for execution environments based on
a ``pay per use'' pricing policy. Conservative synchronization is in
general communication-bound and does not make effective use of CPUs.
On the other hand, in a optimistic simulation, a very large part of
the computation can be thrown away due to roll-backs. Very often the
roll-backs do not remain clustered in a part of the simulator as they
spread to the whole distributed execution architecture. This diffusion
is usually implemented using specially crafted messages (called
anti-messages) that consume a non-negligible amount of
bandwidth. Finally, to implement the roll-back support mechanism it is
necessary to over-provision many parts of the execution architecture
(e.g.,~volatile memory). In a Public Cloud environment, all these
aspects can have a very large impact on the final cost paid for
running the simulation.

To summarize, the Public Cloud comes with a new cost model.  Up to
now most of the budget was for the hardware, the simulation software
tools and writing the simulation model. Now it is no longer necessary
to buy hardware for running the simulations, as computation is one of
the many services that can be rented.  If the goal is to have
simulations that really follow the ``everything as a service''
paradigm, then we need some better way for building PADS, with
mechanisms that need to be less ``expensive'', both in terms of
computation and communication requirements. In this case, the main
evaluation metric should not be the execution speed but the execution cost
(or more likely a combination of both).
A deeper look at the cost model of Public Clouds
(e.g.~Amazon~\cite{amazonEC2pricing}) reveals some interesting
facts. The pricing is often complex, with many
different choices, configurations and options. For example, a new
customer willing to implement a new service on top of the Amazon EC2
Cloud has to choose among several options: ``On-Demand
Instances\footnote{In the Cloud Computing terminology, an instance is
a virtual machine running some operating system and application
software. A new instance can usually be obtained and booted in some
minutes or less.}'', ``Reserved Instances'' (Light, Medium or Heavy
Utilization versions) and ``Spot Instances''. Furthermore, there are
many instance types to choose from (e.g.,~General Purpose, Compute
Optimized, GPU, Memory Optimized, Storage Optimized, Micro), each of
them with many available options. Finally, for the dismay of many
costumers, the price of each instance changes in the different regions
(i.e.,~zones of the world). It is clear that each price and the
detailed service specifications (e.g.,~the amount of free data
transfer provided to each virtual instance) are part of a business
strategy and therefore can change very quickly. Focusing more on the
technical aspects, many of the ``elastic'' features (the possibility
to dynamically scale up or down the resources (e.g.~computation,
memory, storage) provided by the Cloud have been built specifically
for Web applications and therefore cannot be easily used for building
simulators.

In~\cite{Vanmechelen2013126}, the authors analyze the cost optimization
of parallel/distributed simulations run on the Amazon EC2 platform.
More in detail, a cost to performance metric is defined and it used to 
find what EC2 instance type delivers the highest performance per dollar.
In the paper, the most common EC2 instance types are analyzed
under some assumptions (e.g.~it is not considered that each partial 
instance-hour consumed is billed as a full hour). More in detail,
the authors find that, under their metric, the best strategy is to use
large (and costly) instances. This happens because in this kind of instance
it is possible to cluster together many LPs and therefore minimize the 
network load.

\subsection{Performance}\label{sec:performance}

We now focus our attention on the performance of a PADS implemented on
a Public Cloud, when then goal is to obtain the results as fast as
possible (i.e.~the minimizing the WCT).

As usual, even if there are many aspects that could be considered, we
focus on synchronization. Limiting the discussion on synchronization 
is not a concern given that if synchronization has poor performances 
then the rest of the simulator will not do much better. What would 
happen if the synchronization algorithms described in 
Section~\ref{Synchronization} are run on a Public Cloud without 
any modification? What performance can be expected?

Our analysis starts with the simplest synchronization algorithm: 
the time-stepped. As said before, the simulation time is divided in 
a sequence of steps and it is possible to proceed to the next timestep 
only when all the LPs have completed the current one. It is clear that 
the execution speed is bounded by the slowest component. This can be very 
dangerous in execution environments in which the performance variability 
is quite high. The whole simulation could have to stop due to a single LP 
that is slow in responding. This could happen for an imbalance in the model 
partitioning or for some network issues. 

What about the Chandy-Misra-Bryant algorithm? We have already said that 
this algorithm is very demanding in terms of communication resources and 
that, in this case too, a slow LP could be the bottleneck of the whole 
simulation. The last possibility is to use an optimistic synchronization 
algorithm such as the Jefferson's timewarp~\cite{timewarp}. This algorithm 
is not very promising either: timewarp is well-known to have very good 
performance when all LPs can proceed with an execution speed that is 
almost the same. This usually means that all LPs have to be very 
homogeneous in terms of hardware, network performance and load. 
Otherwise, the whole simulation would be slowed down by the 
roll-backs caused by the slow LPs. A requirement that is hard to satisfy 
in a Public Cloud environment. All these synchronization approaches 
have some characteristics that do not fit well with the Public 
Cloud architecture. A more detailed analysis shows that each approach 
has some pros and cons but the key problem is that all of them are 
not adaptable. It is just like working on dynamic problems with a 
static methodology.

In~\cite{Vanmechelen2013126}, the authors assess the performance and 
cost efficiency of different conservative time synchronization protocols 
(i.e.~null-message sending strategies) when a distributed simulation 
is run on a range of Cloud resource types that are available on Amazon 
EC2 (i.e.,~different instance types). This work demonstrates that the 
simulation execution time can be significantly reduced using 
synchronization algorithms that are tailored for this specific 
execution environment and furthermore it shows that the performance 
variability, which is typical in low price instance types, has a 
noticeable impact on performance. Moreover, the authors foresee 
the adoption of dynamic forms of partitioning of the simulation 
model.

An interesting aspect of Cloud Computing is that it allows the
application to shape the underlying execution infrastructure, rather
than the other way around. This means that the application may require
more resources (e.g., instantiate other computing nodes) or release
them at run time. Focusing on computing power, a Cloud application may
require a larger/lower number of computing nodes of the same type of
those already running (\emph{horizontal scaling}), or request to
upgrade some or all the current computing nodes to a higher
configuration, e.g., by asking for a faster processor or more memory
(\emph{vertical scaling}).

PADS could benefit from the above scaling opportunities, since they
could be used to overcome the load balancing issues. Specifically, a
simulator could request more computing nodes to reduce the granularity
of the model partitioning (horizontal scaling), with the goal of
reducing the CPU utilization of processor-bound simulations. On the
other hand, for communication-bound models the simulator could
consolidate highly interacting LPs on the same processor to replace
remote communications with local ones. In this case, the simulator may
request the Cloud to upgrade the nodes where the higher number of LPs
are located (vertical scaling) in order to avoid the introduction of a
CPU-bound bottleneck.

Deciding \emph{when} and \emph{how} to scale is still an open
  problem. For the \emph{when} part, the user needs to define a
  decision procedure that tells when additional resources must be
  requested (\emph{upscaling}) or when some of the resources being
  used can be relinquished to the Cloud provider
  (\emph{downscaling}). Upscaling may be required to speed up the
  simulation, e.g., to meet some deadline on the completion time;
  upscaling may also be used during interactive simulations to focus
  on some interesting phenomena with a higher level of
  detail. Upscaling brings the issue of \emph{how} to scale, i.e.,
  choosing between horizontal and vertical scaling. The decision
  ultimately depends on the cost model being used (refer again to
  Section~\ref{sec:cost-model}). If the user is only interested in
  maximizing the performance of PADS, then vertical scaling is more
  likely to provide benefits since it consolidates the workload on a
  lower number of more powerful nodes, thus reducing the impact of
  communication costs. If the cost model takes into consideration both
  performance and cost, then the choice between vertical and
  horizontal scaling becomes less obvious.

\section{The Quest for Adaptivity}\label{sec:artisgaia}

So far, we have analyzed some of the limitations of current PADS
approaches. Before attempting to address these limitations, it is
important to realize that there is no ``silver bullet'', i.e., no
single solution that addresses them all in a comprehensive and
coherent way. The last attempt to obtain a ``one size fits all''
approach to PADS resulted in the the IEEE 1516
standard~\cite{ieee1516} that, as we mentioned, attracted several
criticisms due to its complexity and
limitations~\cite{Davis:1999:HLA:324898.325337,hlainteroperability,4736178,caihlaperformance}.

In this section we describe an approach to build scalable and adaptive
simulation models by addressing the \emph{partitioning
  problem}~\cite{gda-ijspm-2009}, that is the decomposition of the
simulation model into a number of components and their allocation
among the execution units. Our approach aims at achieving two goals:
balance the computation load in the execution architecture and
minimize the communication overhead~\cite{bagrodia98}. If both these
requirements are satisfied, then the simulation execution is likely to
be carried out efficiently. The difficult part is that the balancing
procedure should be transparent to users and adaptive.  Adaptivity is
of extreme importance given that both the behavior of the simulation
model and the state of the underlying execution architecture can not
be accurately predicted.

In our view, the adaptive partitioning of the simulation model is a 
prerequisite for a solution to most of the PADS problems described in the 
previous sections.

\subsection{Model Decomposition}

A complex simulation model should be partitioned in smaller parts
referred to as Simulated Entities (SEs). Each SE interacts through
message exchanges with other SEs to implement the expected
behavior. We assume that the execution environment consists of a set
of interconnected Physical Execution Units (PEUs).  For example, each
PEU can be a core in a modern multi-core CPU, a processor in a shared
memory multiprocessor, a node in a LAN-based cluster or even a Cloud
instance. Following the PADS approach, the simulated model is
partitioned among all the PEUs and each PEU is responsible for the
execution of only a part of the model. In a traditional PADS, the
model is partitioned in a set of Logical Processes (LPs) and each LP
runs on a different PEU. Instead, in our case the LPs act as
containers of SEs. In other words, the simulation model is partitioned
in basic components (i.e., the SEs) that are allocated within the
LPs. The SEs does not need to be statically allocated on a specific
LP; indeed, each SE can migrate to improve the runtime efficiency of
the simulator~\cite{gda-dsrt-2004}, as will be described shortly. For
better scalability, new LPs can be allocated during the simulation,
and idle ones can be disposed.

In practice, the simulation is organized as a Multi Agent System
(MAS)~\cite{Wooldridge:2009:IMS:1695886}. The MAS paradigm has the
potential to enhance usability and transparency of simulation tools
(as discussed in Section~\ref{sec:challenges.transparency}). Choosing
the proper granularity of the SEs can be problematic. Having a large
number of very simple entities will probably increase the
communication and management overhead. Conversely, having only a few
big entities means that the workload can be spread less effectively
across the available PEUs, resulting in bad load balancing. The
appropriate granularity of entities can sometimes be ``suggested'' by
the simulation model itself; for example, it is pretty natural to
model each node in a wireless network as a single SE.

\subsection{Dynamic Partitioning}

To properly partition the model, we have to consider two main
aspects. First, a PADS has to deal with a significant communication
cost due to network latency and bandwidth limitations. Second, the
execution speed of the simulator is bounded by its slowest component
and therefore effective load balancing strategies must be implemented
to quickly identify and remove performance ``hot spots''. As seen in
Section~\ref{sec:performance}, both the communication overhead and the
computation bottlenecks have a big influence on the simulators 
performance (e.g.,~synchronization).

An effective strategy to reduce communication costs is to cluster the
strongly interacting SEs together within the same
LP~\cite{gda-pads-2003}. However, a too aggressive clustering may
result in poor load balancing if too many SEs are brought inside the
same LP.  Since communication patterns may change over time, the
optimal partitioning is in general time-dependent, and is the result
of a dynamic optimization problem with multiple conflicting goals and
unknown, time-varying parameters.

\begin{figure}[ht]
\centering
\subfloat[\label{fig:structure}]{\includegraphics[width=9.0cm]{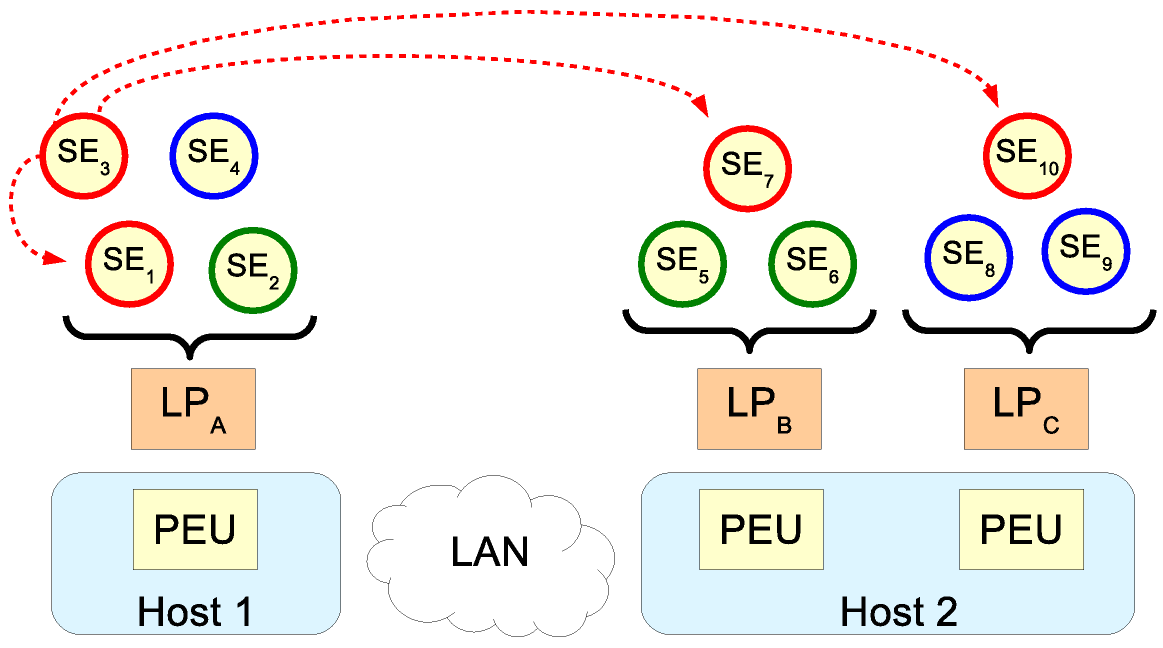}}\\
\subfloat[\label{fig:migration}]{\includegraphics[width=9.0cm]{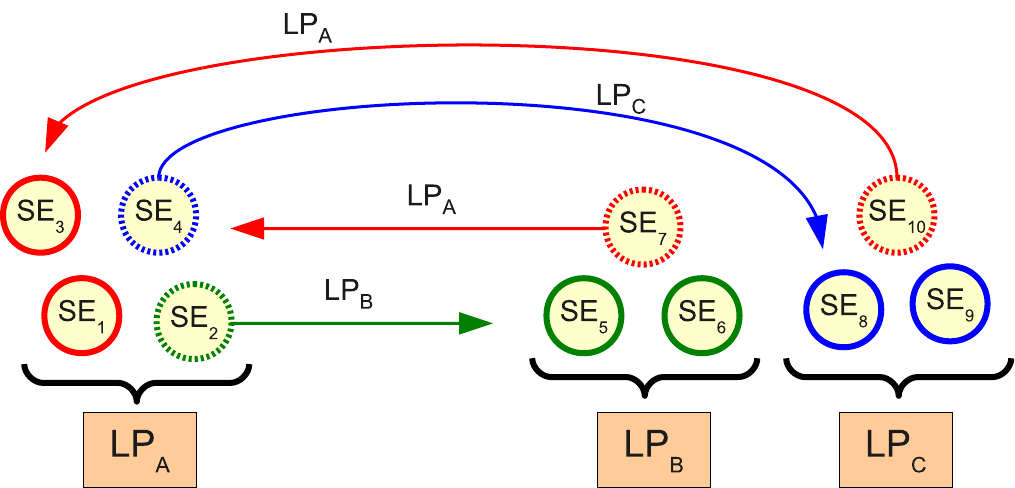}}
\caption{A PADS with three LPs and ten SEs. \ref{fig:structure}) Initial situation; \ref{fig:migration}) After migration.}
\end{figure}

Figure~\ref{fig:structure} shows an example where a distributed
simulation runs on two hosts connected by a local area network. Host 1
has a single PEU (e.g.,~a processor with a single core) and executes
one LP, Host 2 has two PEUs executing one LP each. An LP is the
container of a set of SEs that interact through message exchanges
(depicted as dotted lines in the figure).  The colors used to draw the
SEs represent the different interaction groups, that is, groups of SEs
that interact the most.  The interaction groups are $\{\mathrm{SE}_1,
\mathrm{SE}_3, \mathrm{SE}_7, \mathrm{SE}_{10}\}$, $\{\mathrm{SE}_2,
\mathrm{SE}_5, \mathrm{SE}_6\}$ and $\{\mathrm{SE}_4, \mathrm{SE}_8,
\mathrm{SE}_9\}$.  If these interaction groups are expected to last
for some amount of time, a more efficient allocation that reduces the
communication overhead would be to migrate SEs as shown in
Figure~\ref{fig:migration}, so that each interaction group lies within
the same LP. In this simple example, interaction groups are of
approximately the same size, and therefore the new allocation is well
balanced.

Since the interaction pattern between SEs may change unpredictably, it
is necessary to rely on heuristics that monitor the communication and
the load of each LP, and decide if and when a migration should
happen. Obviously migrations have a non-negligible cost that includes
the serialization of state variables of the SE to be migrated, the
network transfer delay, and the de-serialization step required before
normal operations can be restored at the destination LP.

The overall execution speed of the whole simulation is therefore the
result of two competing forces: one that tries to aggregate SEs to
reduce communication costs, and the other one that tries to migrate
SEs away from overloaded LPs. Our experience with the practical
implementation of adaptive migration strategies is reported in the
following.

\subsection{ART\`IS and GAIA+}

\begin{figure}[ht]
\centering
\includegraphics[width=9.0cm]{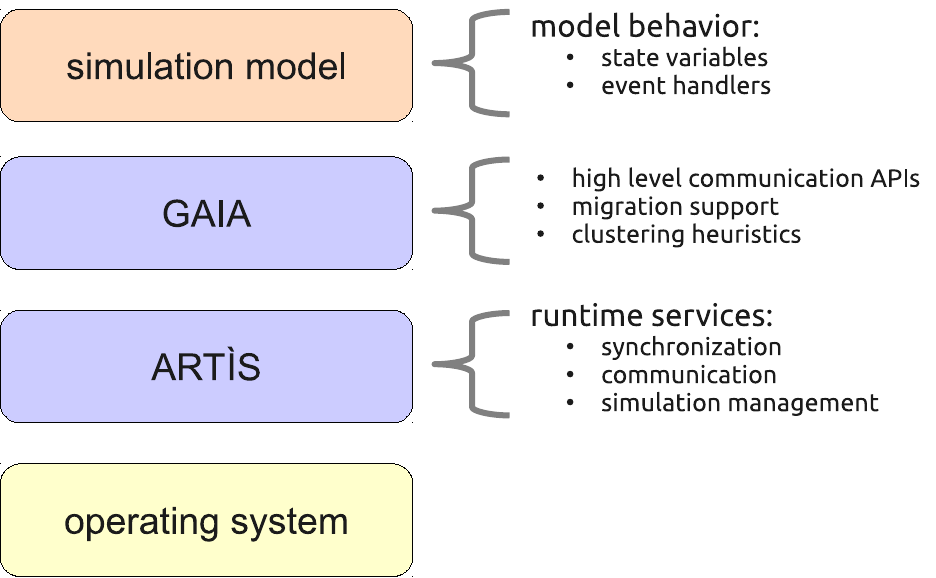}
\caption{High level structure of the simulator and its main components}
\label{fig_logicscheme}
\end{figure}

In the last years, we pursued the approach described so far with the
realization of a new simulation middleware called Advanced RTI System
(ART\`IS) and the companion GAIA+ framework (Generic Adaptive
Interaction
Architecture)~\cite{gda-ijspm-2009,gda-dsrt-2004,gda-pads-2003}. The
high level structure of our simulator is shown in
Figure~\ref{fig_logicscheme}. The upper layer (labeled
\emph{Simulation Model}) is responsible for the allocation of the
state variables to represent the evolution of the modeled system, the
implementation of the model behavior and the necessary event
handlers. The simulation model is built on top of the GAIA+ software
framework, that provides the simulation developer with a set of
services such as high level communication primitives between SEs,
creation of SE instances and so on. GAIA+ can analyze the interactions
(i.e., message exchanges) between SEs to decide if and how SEs should
be clustered together, also providing the necessary support
services. Given the distributed nature of a PADS, each LP analyzes the
interactions of all SEs it hosts, and takes migration decisions based
on local information and information received from other LPs, so that
adaptivity can be achieved without resorting to any centralized
component. Finally, in the layered architecture shown in the figure,
the ART\`IS middleware~\cite{pads} provides common PADS
functionalities such as synchronization, low level communication
between LPs, and simulation management services.

To validate ART\`IS and GAIA+, a number of models have been
implemented, including wired and wireless communication
environments~\cite{gda-simutools-09,moves}: using the ideas previously
described, it has been possible to manage the fine grained simulation
of complex communication protocols such as~IEEE 802.11 in presence of
a huge number of nodes (up to one million)~\cite{gda-ijspm-2009}. In
the wired case, we are working on the design and evaluation of gossip
protocols in unstructured networks (e.g.,~scale-free, small-world,
random)~\cite{gda-disio-10,gda-disio-11}. Good scalability has been
achieved, using off-the-shelf hardware, thanks to the use of adaptive
migration and load balancing techniques. Both this software tools have
been designed to be platform-independent. In particular, the dynamic
partitioning feature provided by GAIA+ fits very well with the
horizontal and vertical scalability provided by Cloud Computing.

The ART\`IS middleware, the GAIA+ framework, sample simulation models,
and the support scripts and scenario definition files are freely
available for research purposes~\cite{pads}. A large part of the
software is provided in both binary and source form. We expect to make
the source code for all components available through an open source
license.

\subsection{Case study}

A complete evaluation of GAIA+/ART\`IS is outside the scope of this
paper; however, a simple case study can be useful to better understand
the proposed mechanism, and will be illustrated in this section. For
the sake of simplicity, we consider only the adaptive clustering
approach used to reduce the communication overhead. Advanced forms of
load-balancing and reaction to background load have been implemented
as well, but will not be examined here.

We consider a time-stepped, ad hoc wireless network model of $9999$
Mobile Hosts (MHs). The simulated scenario is a two-dimensional area
($10000 \times 10000$ units) with periodic boundary conditions, in
which each MH follows a random waypoint mobility model ($\mathit{max
  speed}=10$ space-units/time-unit, $70\%$ of nodes move at each
timestep). The communication model is very simple and does not
consider the details of low level medium access control. At each
timestep, a random subset of $20\%$ of the MHs broadcasts a ping
message to all nodes that are within the transmission radius of $250$
space units. This model captures both the dynamicity of the simulated
systems and the space locality aspects of wireless communication. The
MHs have been partitioned in $3$ PEUs, each one running a single LP.

\begin{figure}[!t]
\centering
\subfloat[t=0]{\includegraphics[width=6.3cm,angle=270]{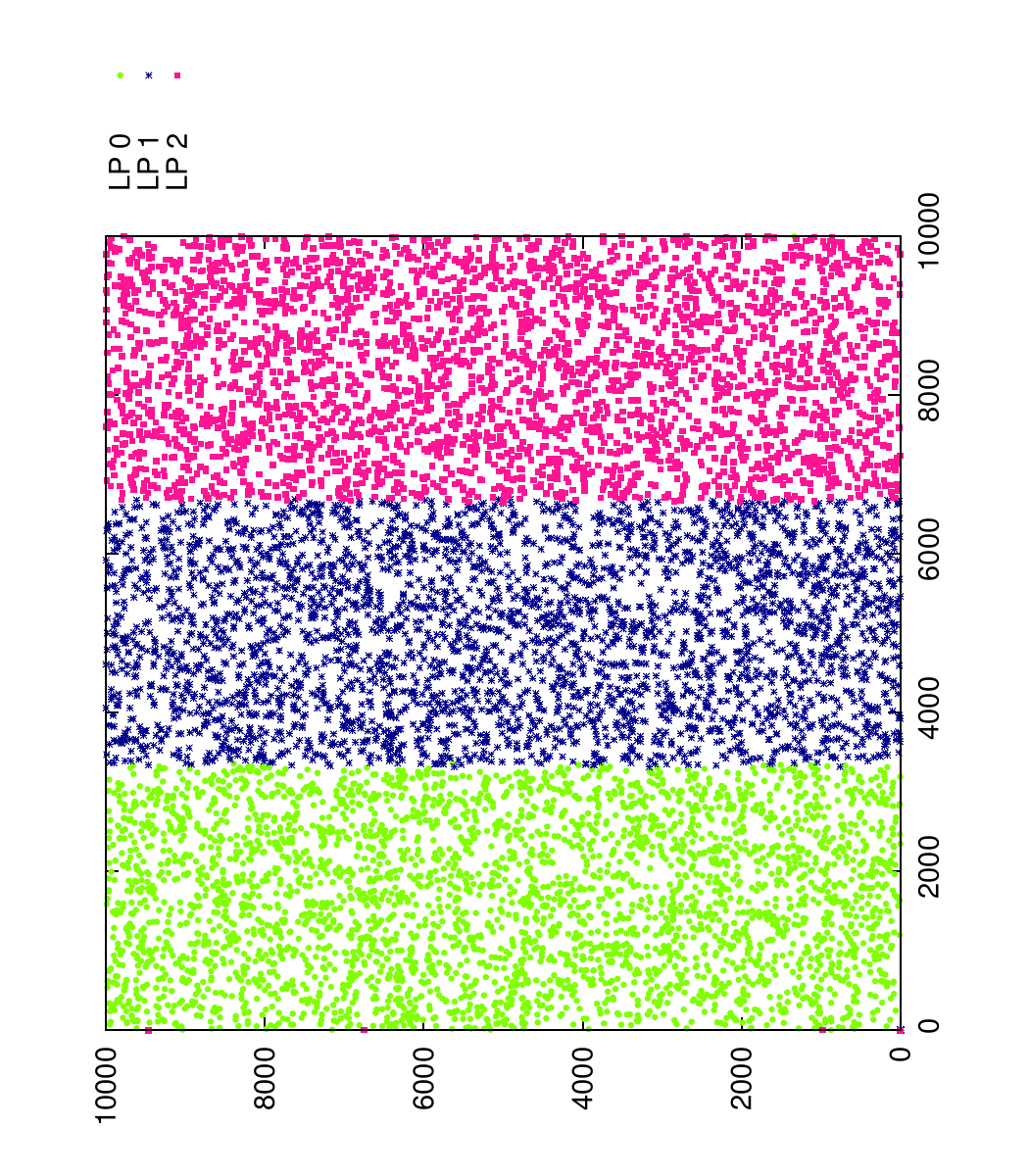}\label{fig_nogaia:a}}
\subfloat[t=333]{\includegraphics[width=6.3cm,angle=270]{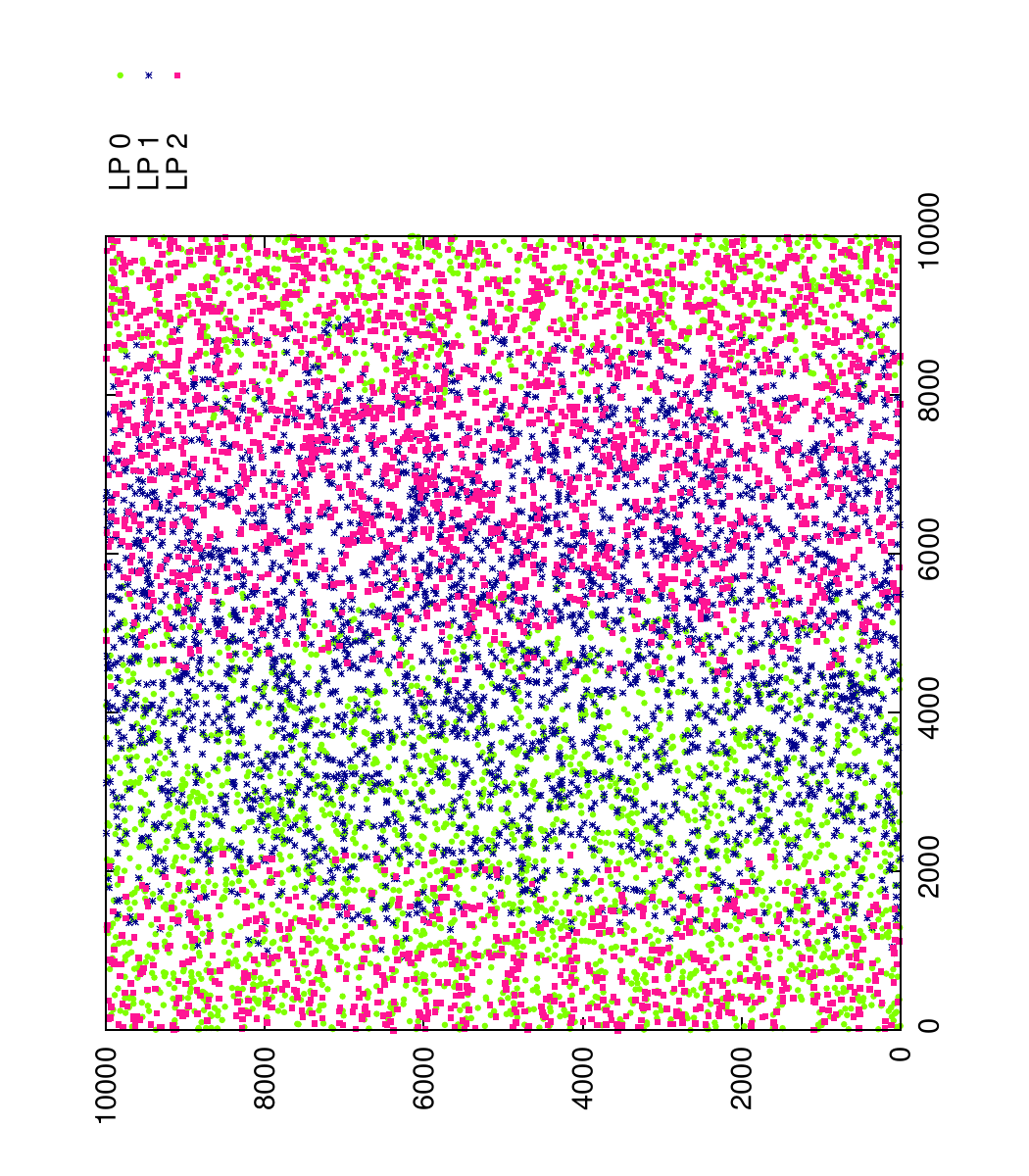}\label{fig_nogaia:b}}\\
\subfloat[t=666]{\includegraphics[width=6.3cm,angle=270]{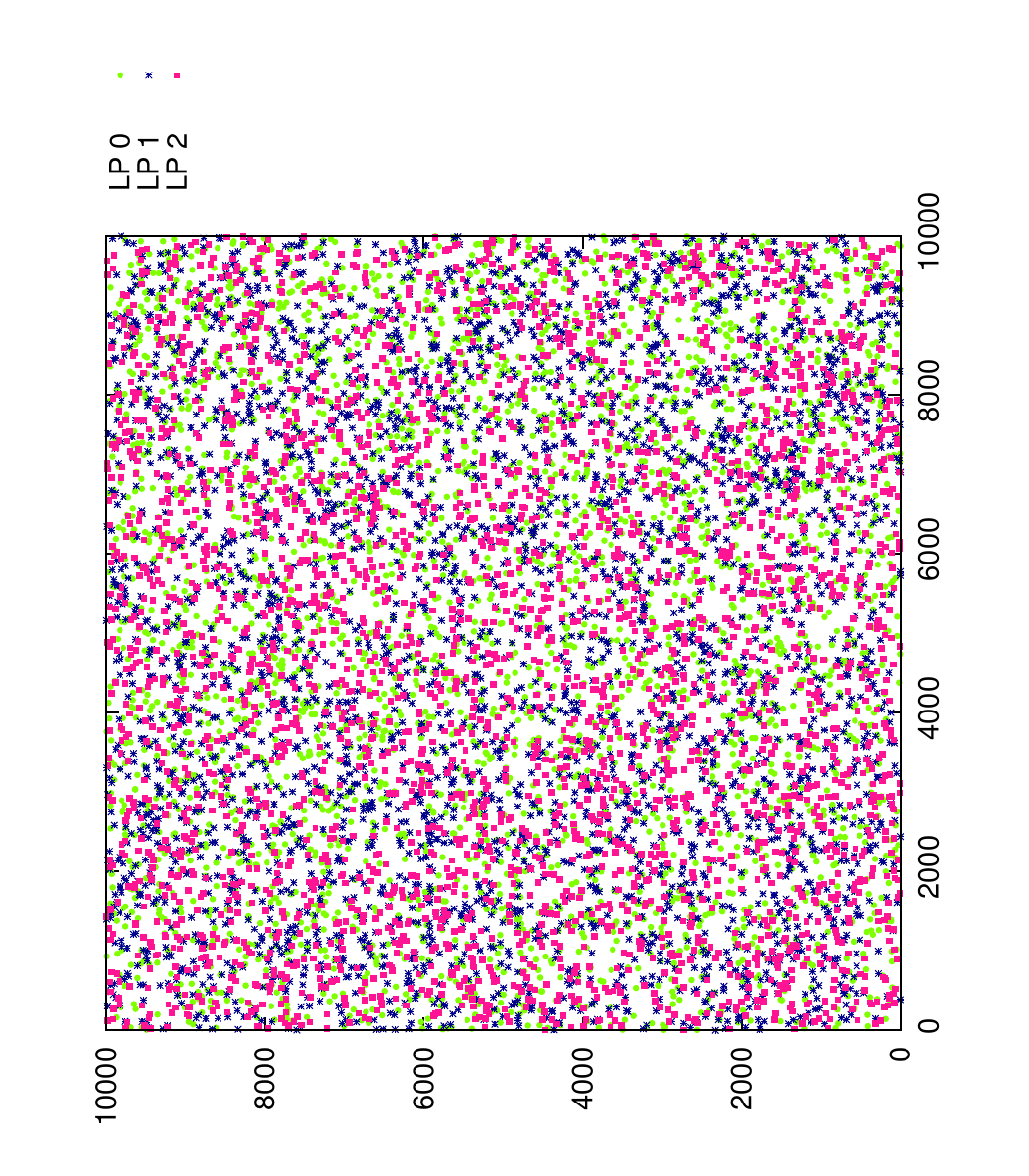}\label{fig_nogaia:c}}
\subfloat[t=999]{\includegraphics[width=6.3cm,angle=270]{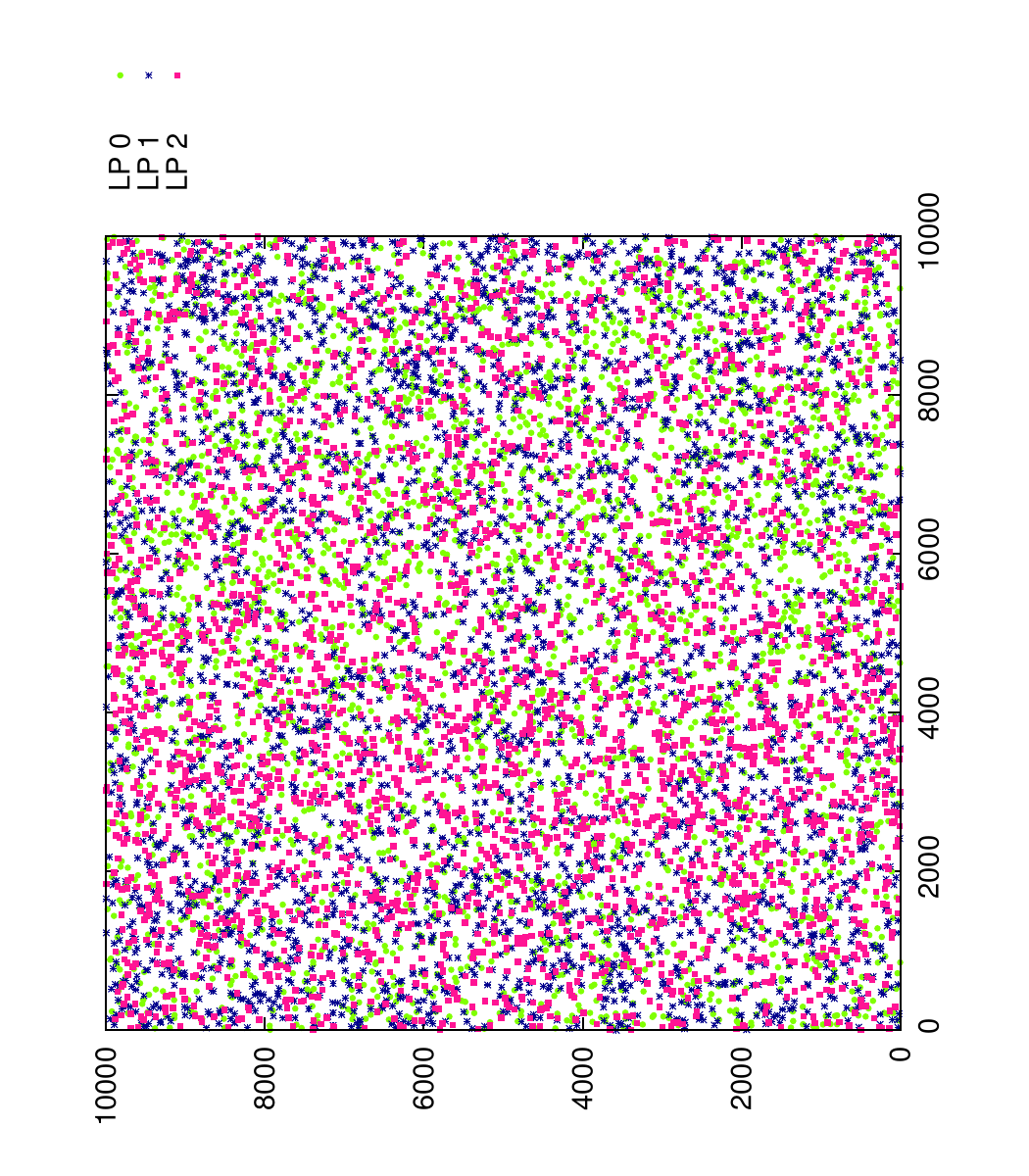}\label{fig_nogaia:d}}
\caption{Spatial position and allocation in LPs of the mobile wireless
  nodes simulation at different time steps. The color and shape of
  dots shows in which LP each node is allocated. Adaptive migration is
  not active (GAIA+ off).}
\label{fig_nogaia}
\end{figure}

Our analysis starts with a scenario in which simulated entities do not
migrate across LPs (the GAIA+ migration engine is turned off).  To
balance the workload evenly across the three LPs, each one should
receive approximately $9999 / 3 = 3333$ MHs.  A simple metric that can
be used to assess the ability to properly cluster the simulated
components is the Local Communication Ratio (LCR), defined as the
percentage of local messages with respect to the total messages sent
or received by a simulated component (higher is better). A random
assignment of MHs to LPs results in a LCR of $\dfrac{100}{\#LP}$ (\%),
where $\#LP$ as the number of LPs in the PADS. By taking into
consideration the model semantics, we know that the communication
between entities exhibits a strong spatial locality within the
simulation area, since each MH only interacts with a local
neighborhood. Therefore, a better assignment of MHs to LPs can be
obtained by partitioning the simulated area, e.g., in vertical
stripes, and to allocate all MHs in each stripe to the same LP, as
shown in Figure~\ref{fig_nogaia:a}. However, such allocation works
well for the initial simulation steps, but then quickly degenerates as
the spatial position of MHs change as the results of their movement.
The situation can be observed in Figures~\ref{fig_nogaia:b}
through~\ref{fig_nogaia:d}, where the initial spatial locality of MHs
is rapidly destroyed.

\begin{figure}[ht]
\centering
\subfloat[t=0]{\includegraphics[width=6.3cm,angle=270]{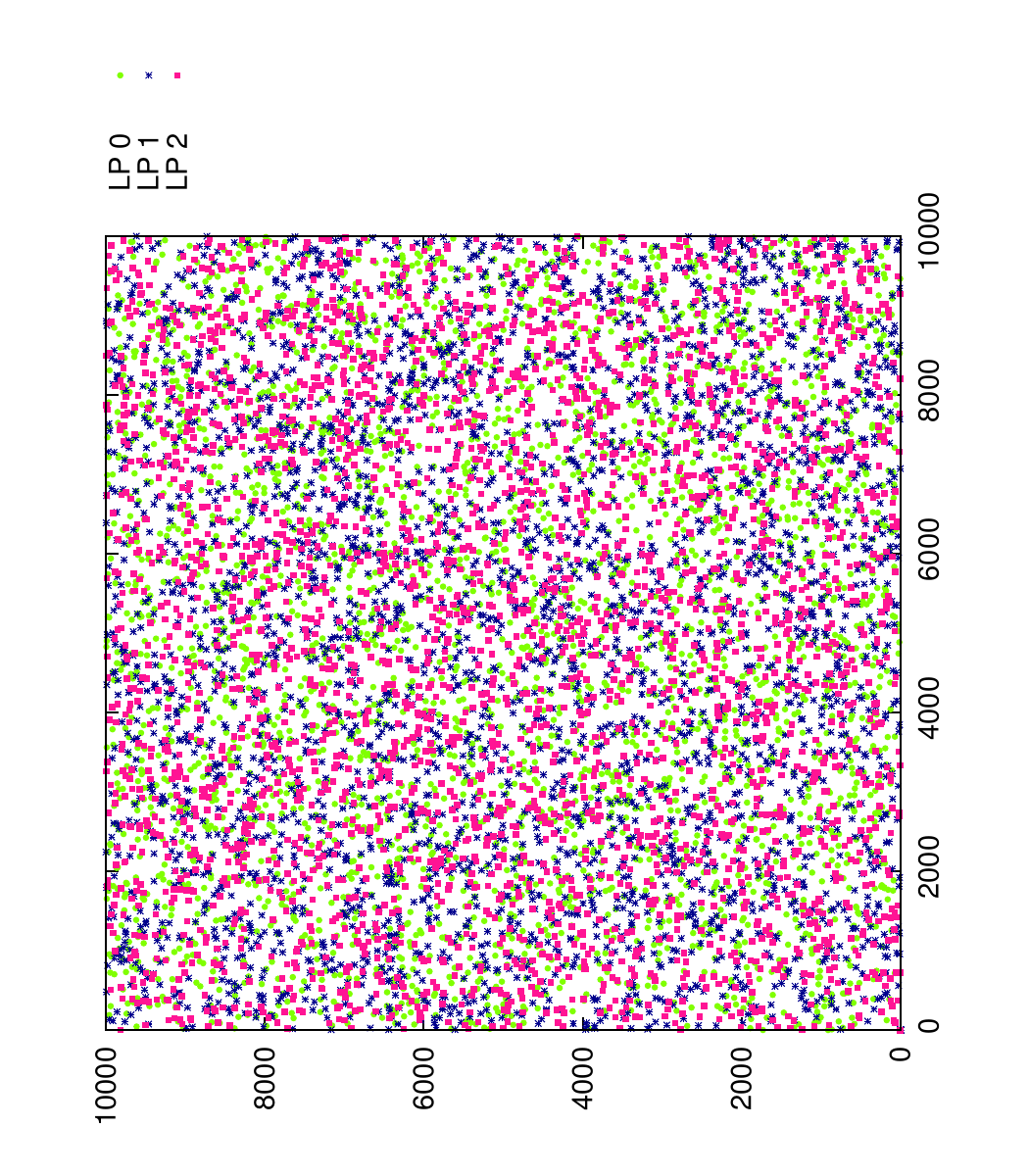}\label{fig_initial_and_final:a}}
\subfloat[t=999]{\includegraphics[width=6.3cm,angle=270]{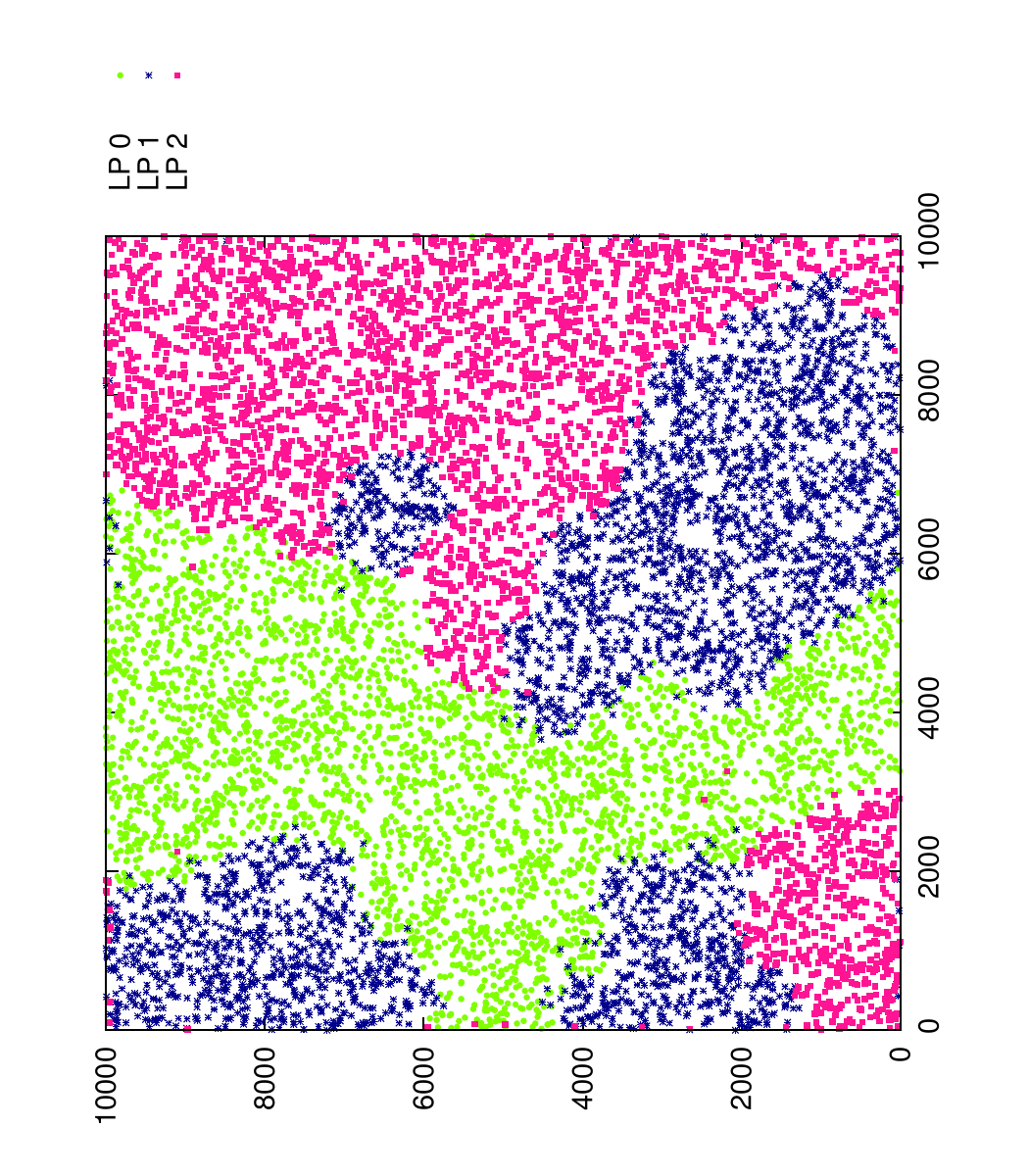}\label{fig_initial_and_final:b}}
\caption{Spatial position and allocation in LPs of the mobile wireless
  nodes at different time steps. The color and shape of dots shows in
  which LP each node is allocated. Adaptive migration is activated
  (GAIA+ ON).}
\label{fig_initial_and_final}
\end{figure}

Figure~\ref{fig_initial_and_final} shows what happens when the
adaptive migration facility of GAIA+ is turned on, starting from a
completely random allocation of MHs to LPs.  GAIA+ analyzes the
communication pattern of each MH and clusters the interacting hosts in
the same LP. A snapshot at the end of the run
(Figure~\ref{fig_initial_and_final:b}) shows that the MHs have been
clustered in groups, and that each group is determined by its position
in the simulated area. It is interesting to observe that
Figure~\ref{fig_initial_and_final:b} is similar to what would be
produced by Schelling's segregation model~\cite{segregation}; indeed,
each MH has a preference towards the other MHs with which it
communicates the most, and GAIA+ tries to cluster those MHs together
within the same LP.

\begin{figure}[ht]
\centering
\subfloat[Random initial allocation]{\includegraphics[width=4.6cm,angle=270]{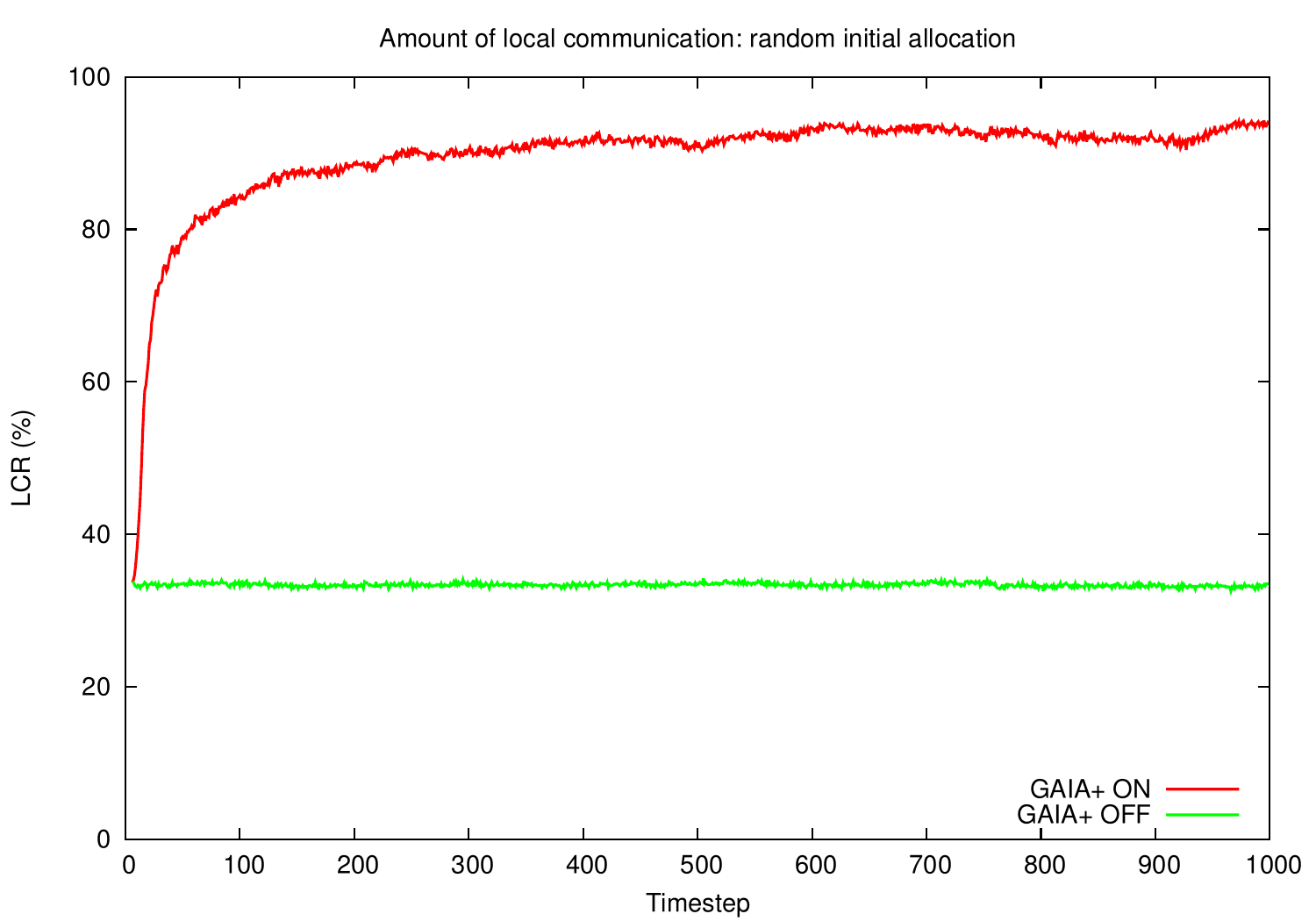}\label{fig:lcr_a}}
\subfloat[Initial allocation in stripes]{\includegraphics[width=4.6cm,angle=270]{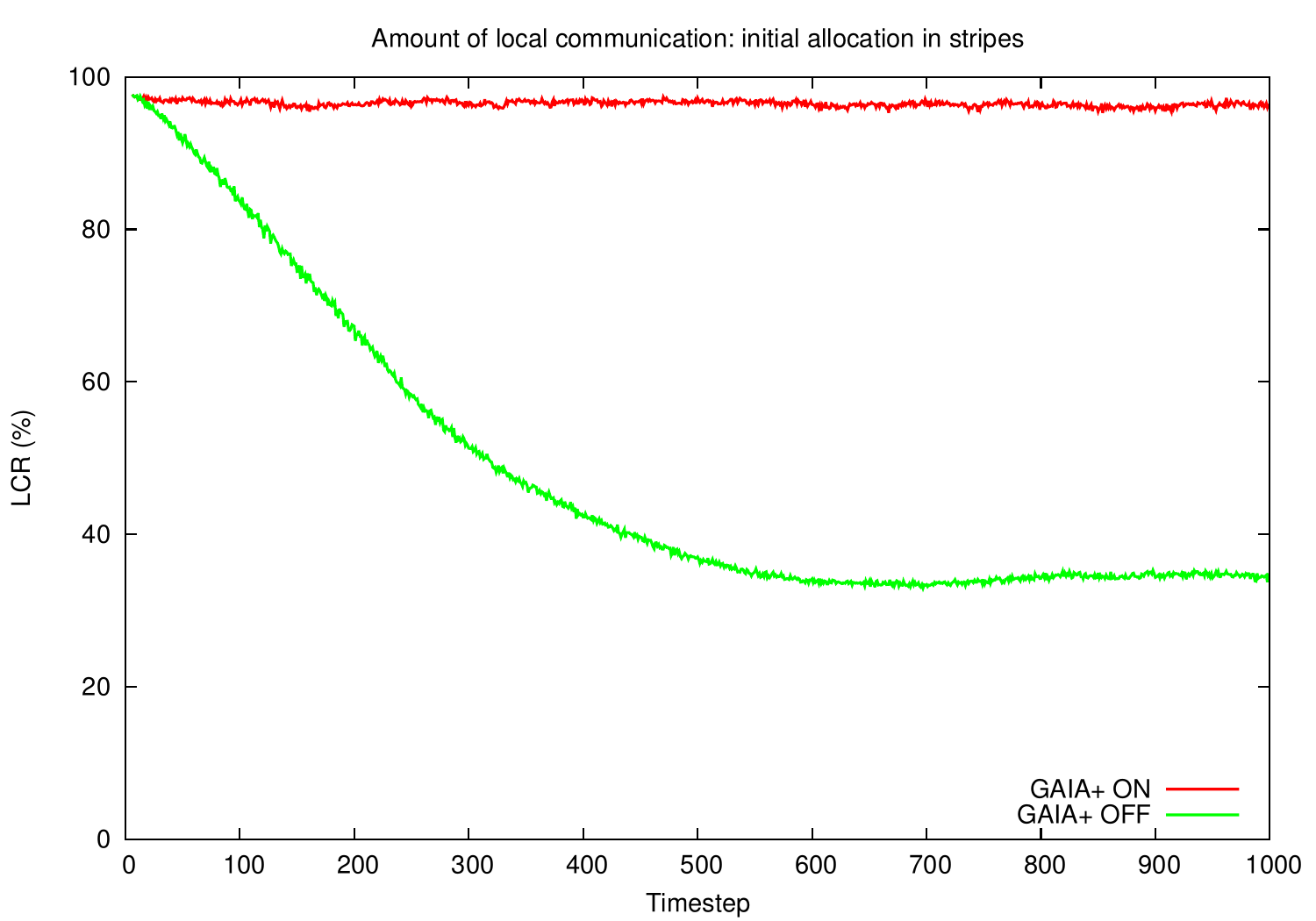}\label{fig:lcr_b}}
\caption{Local Communication Ratio (LCR) evolution with different
  initial allocations.}
\label{fig:lcr}
\end{figure}

The performance of the static allocation versus the dynamic allocation
implemented by GAIA+ is shown in Figure~\ref{fig:lcr}. The green line
shows the mean LCR as a function of the simulated time step if MHs are
statically assigned to the LPs at the beginning, while the red line
shows the mean LCR when the adaptive migration of GAIA+ is
enabled. Figure~\ref{fig:lcr_a} corresponds to the situation in which
all MHs are initially randomly assigned to the LPs. As discussed
above, the expected LCR in this situation is about $33\%$, which is
the initial value at step $t=0$. As the simulation proceeds, the LCR
provided by the static allocation remains at the initial value.  On
the other hand, if the adaptive migration feature (GAIA+) is activated
then the LCR rapidly increases.

In Figure~\ref{fig:lcr_b} we use the ``sliced'' allocation at $t=0$,
where MHs are assigned to LPs according to their initial position.  In
the case of static mapping (green line), the LCR drops to $33\%$ as
the movement of MHs destroys the initial statically imposed
locality. If adaptive migration is used, the LCR remains stable at a
nearly optimal value.

It is important to remark that a high LCR value does not guarantee, by
itself, that a PADS scales well. In fact, migrating simulation
entities does have a cost that can be significant if a large amount of
state information needs to be moved from one PEU to another. However,
previous studies have shown that good scalability can indeed be
achieved even for models that do not exhibit the simple spatial
locality of wireless ad-hoc networks~\cite{gda-mospas-11}.

It is important to remark that the particular metric considered in
this section (LCR) is independent from the execution environment
(e.g.,~private clusters, Private or Public Clouds). In fact, a high
LCR value demonstrates that the dynamic clustering of the simulated
entities performed by GAIA+ leads to a reduction in the communication
overhead since most communications are local. The actual impact of
this reduction on the simulation wall-clock time depends on the
(absolute) communication costs of the specific execution environment.

\subsection{Fault Tolerance}

So far, we have described our effort in building adaptive PADS. In our
opinion, this is a first step towards simulations that are able to run
efficiently on modern computing infrastructures. A key point is still
missing: the support for fault tolerance in distributed
simulation. Usually, if a LP crashes (e.g.,~due to hardware failure of
the underlying PEU), then the whole simulation aborts. In the case of
a long-running simulation that is executed on hundreds or thousands of
PEUs, the probability of a hardware failure during execution can not
be neglected. Therefore, fault tolerance is essential, especially if
we wish to run large-scale simulations on cheap but unreliable
execution systems such as resources provided by a Cloud
infrastructure.

A new software layer called GAIA Fault-Tolerance (GAIA-FT) extends the
interface provided by GAIA+ with provisions for fault tolerant
execution. A GAIA-FT simulation is made by a set of interacting
Virtual Simulated Entities (VSEs). A VSE is implemented as a set of
identical SEs running on different LPs, therefore achieving fault
tolerance through replication.

By selecting the number of replicas it is possible to decide how many
crashes we want to tolerate, and also cope with byzantine
failures. From the user's point of view, a VSE is just a special type
of SE; in other words, the replication introduced by GAIA-FT is
totally transparent and is built on the lower layers of the GAIA+
architecture. This means that the adaptive strategies described so far
are still available and can be used to migrate replicas within a VSE
to improve load balancing and communication. The basic implementation
of GAIA-FT has been recently completed, and we are in the process of
evaluating the replication mechanism.

Of course, fault tolerance comes at a cost, since replication
increases the number of SEs to be partitioned. Therefore requiring
more resources to execute the simulation, both in term of execution
nodes and wall-clock time. This means that, a larger number of LPs
will be used and that they will need to be kept synchronized in order
for the simulation to advance, so the issues already described in
Section~\ref{sec:performance} apply. Finally, when multiple copies of
the same SE are created, the simulation middleware must ensure that
all copies reside in LPs that are run on different execution nodes, to
avoid that a failed processor brings down multiple replicas of the
same SE. This makes the load balancing task more complex, since a new
constraint is introduced (namely, multiple copies of the same SE must
never reside on the same LP or on LPs that are run on the same
execution node). Identifying a reasonable trade-off between
reliability and performance in a parallel simulation is an interesting
research topic that we are pursuing.

\section{Conclusions}\label{sec:conc}

In this paper, we have reviewed the main ideas behind Parallel and
Distributed Simulation (PADS), and we observed that current PADS
technologies are unable to fulfill many requirements (e.g.~usability,
adaptivity) in the context of the new parallel computing architectures.

Most computing devices are already equipped with multi-core CPUs, and
Cloud-based technologies, where computation, storage and communication
resources can be acquired ``on demand'' using a pay-as-you go model,
are gaining traction very quickly. The ``simulation as a service''
paradigm has already been proposed as a possible application of Cloud
technologies to enhance simulation tools.

We have shown that, the current PADS techniques are unable to fit well
with these architectures, and that more work is needed to increase
usability and performance of simulators in such conditions. We claim
that a solution for such problems is required to deal with model
partitioning across the execution nodes. To this aim, we propose an
approach based on multi-agent systems. Its main characteristic is the
adaptive migration of the simulated entities between the execution
units. Heuristics can be used to reducing the communication cost and
to achieve better load balancing of the simulation workload. Our
proposal has been implemented in the ART\`IS/GAIA+ simulation
middleware and tested with different models, with promising results.

Finally, the horizontal and vertical scaling possibilities provided by
Cloud systems (see Section~\ref{sec:performance}) deserve further
investigations in the context of PADS. The Cloud Computing paradigm
allows applications to request the type and amount of resources of
their choice at run time, enabling dynamic sizing of the execution
environment. How to enable a larger class of applications to take
advantage of these possibilities is yes to be understood.

\section*{Acknowledgments.} 

We would like to thank the anonymous reviewers whose detailed comments
and suggestions greatly contributed to improve the overall quality of
this paper.

\section*{Acronyms}

\begin{tabbing}
\hspace{15mm}\=\kill
ARTIS \> Advanced RTI System\\
CMB \> Chandy-Misra-Bryant\\
COTS \> Commercial Off-The-Shelf\\
DES \> Discrete Event Simulation\\
EOS \> End of Step\\
GAIA+ \> Generic Adaptive Interaction Architecture\\
GPU \> Graphics Processing Units\\
GVT \> Global Virtual Time\\
HLA \> High Level Architecture\\
HPC \> High Performance Computing\\
HT \> Hyper-Threading\\
LCR \> Local Communication Ratio\\
LP \> Logical Process\\
MAS \> Multi Agent System\\
MH \> Mobile Host\\
PADS \> Parallel and Distributed Simulation\\
PEU \> Physical Execution Unit\\
SE \> Simulated Entity\\
SIMD \> Single Instruction Multiple Data\\
VSE \> Virtual Simulated Entity\\
WCT \> Wall Clock Time
\end{tabbing}

\bibliographystyle{elsarticle-num}
\bibliography{paper}

\end{document}